\documentclass[twocolumn,english,aps,prl,showpacs]{revtex4-1}
\usepackage[T1]{fontenc}
\usepackage{amsmath,graphicx,amssymb,epsfig,babel,dsfont,color}

\begin{document}

\title{Strong slowing down of the thermalization process of solids\\interacting in extreme near-field regime}

\author{M. Reina}

\author{R. Messina}

\author{P. Ben-Abdallah}
\email{pba@institutoptique.fr}
\affiliation{Laboratoire Charles Fabry, UMR 8501, Institut d'Optique, CNRS, Universit\'{e} Paris-Saclay, 2 Avenue Augustin Fresnel, 91127 Palaiseau Cedex, France.}

\date{\today}

\begin{abstract}
When two solids at different temperatures are separated by a vacuum gap they relax toward their equilibrium state by exchanging heat either by radiation, phonon or electron tunneling, depending on their separation distance and on the nature of materials. The interplay between this exchange of energy and its spreading through each solid entirely drives the relaxation dynamics. Here we highlight a significant slowing down of this process in the extreme near-field regime at distances where the heat flux exchanged between the two solids is comparable or even dominates over the flux carried by conduction inside each solid. This mechanism, leading to a strong effective increase of the system thermal inertia, should play an important role in the temporal evolution of thermal state of interacting solids systems at nanometric and subnanometric scales. 
\end{abstract}

\maketitle

The relaxation of bodies in mutal interaction which are initially prepared in two different thermal states is an important problem in physics both on a fundamental~\cite{Evans,Schuster,Prigogine,Jin,Reid,MessinaPRB2013,superdiff1,superdiff2,eigen, phonon} and practical point of view~\cite{Guha,logic,refrige,Yu,Transi,Bhatia}. When these bodies are separated by a vacuum gap this relaxation is mediated by radiative heat exchange or by tunneling of heat carriers (phonons, electrons, excitons...). Usually the system evolves toward a state of equipartition of energy and uniform temperature by maximizing its entropy~\cite{Kubo}. This corresponds to an evolution of all particles inside the system toward the same average energy through various interaction mechanisms even if their initial energies are very different. This evolution is well described for classical systems by the Boltzmnan equation for the probability density of particles, over all their possible states in the phase space.

In this Letter we investigate the relaxation dynamics of systems exchanging heat in a strong-interaction regime. To explore this thermalization process we consider two solids out of thermal equilibrium which exchange heat through a nanometric or subnanometric vacuum gap in the transition domain between the radiative and the conductive regime~\cite{Chen,Reddy,Kittel} also called extreme near-field regime. In this range of separation distance energy exchange competes or even dominates~\cite{Messina2016,Reina2020} with respect to the conductive heat transport within the bodies themselves. In this regime of strong interaction we demonstrate that counterintuitively the relaxation time of the system is dramatically extended compared to a weak-coupling situation and demonstrate that the relaxation time of the system towards thermal equilibrium is dramatically extended.

\begin{figure}
	\centering
	\includegraphics[width=0.4\textwidth]{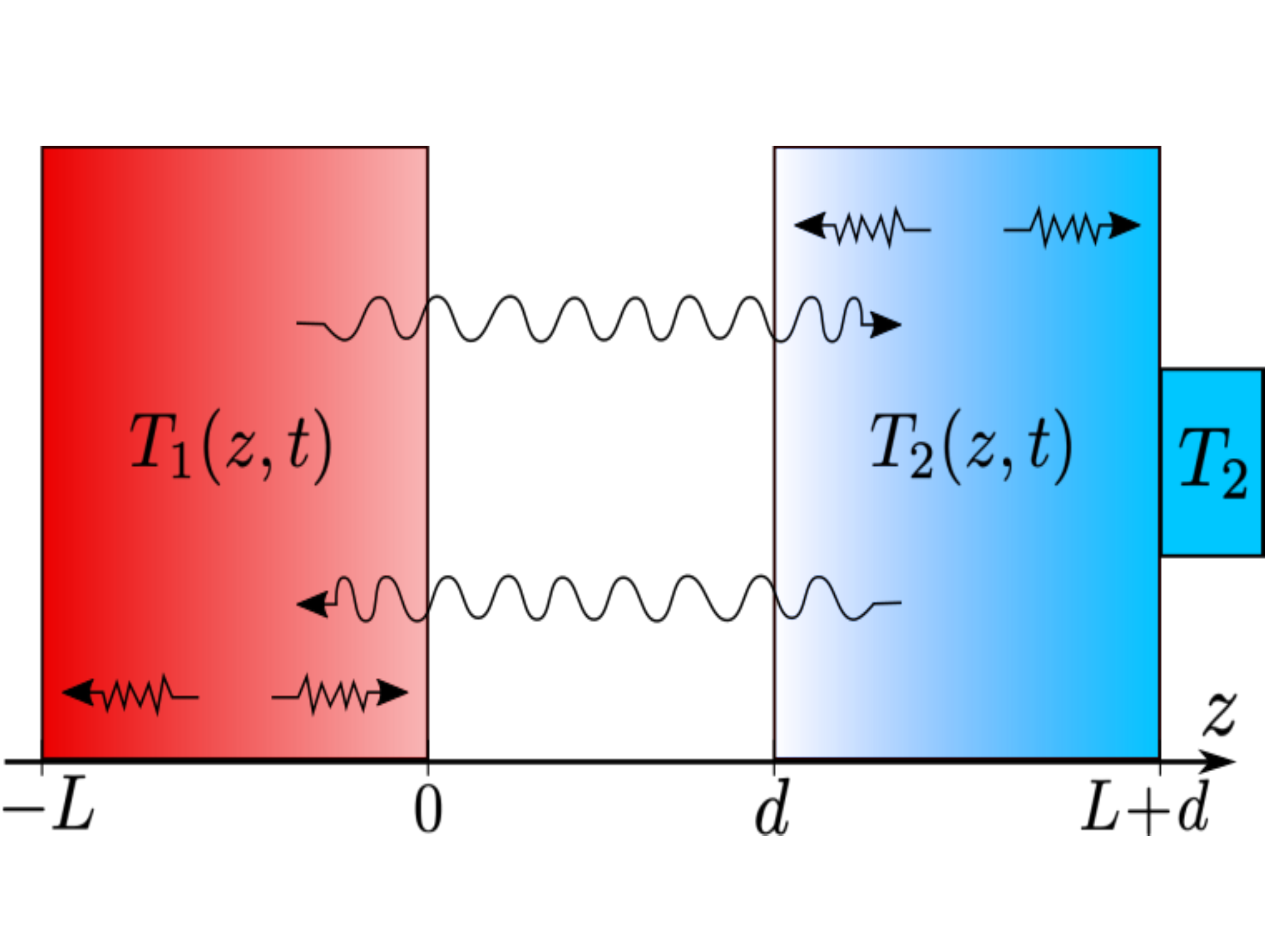}
	\caption{Sketch of system. Two solid films out of thermal equilibrium having temperatures $T_1(z,t)$ and $T_2(z,t)$ are separated by a vacuum gap of thickness $d$. They relax by exchanging heat either by radiation, phonon or electron tunneling depending on the separation distance and the nature of the materials. This transmitted energy is spreaded out through the films thanks to a diffusion process.}
	\label{Fig:Syst}
\end{figure}

To address this problem let us consider two identical solid films as sketched in Fig.~\ref{Fig:Syst}. When the heat transport inside these films is governed by a simple diffusion process the spatio-temporal evolution of the films temperature can be obtained by solving the energy balance equation
\begin{equation}
	\rho C\; \frac{\partial T_k(z,t)}{\partial t}=-\nabla \cdot[\kappa\nabla T_k(z,t)] + P^{j\rightarrow k}(z,t),
	\label{Eq:Coupling}
\end{equation}
where $\rho$ denotes the mass density of the films, $C$ their specific heat capacity, $\kappa$ their thermal conductivity and $P^{j\rightarrow k}(z,t)$ the power density received locally by one film from the opposite one, This power density, as expected, depends implicitly on the entire temperature profiles $T_k(z,t)$ at each time $t$. It is clear that the time evolution of the temperature depends on the interplay between the inner transport mechanism and the heat exchange between the two films. When the flux carried by conduction within the films dominates with respect to the flux exchanged between the two films, the temperature profiles stay uniform at any time within each film and then the two temperatures obey the following equation
\begin{equation}
		\rho C\frac{dT_k}{dt} = P^{j\rightarrow k}[T_j(t),T_k(t)],
		\label{Eq:SystPvH}
\end{equation}
where now $P^{j\rightarrow k}[T_j(t),T_k(t)]$ corresponds to the total power (per unit volume) transferred from film $j$ to film $k$, function of the two temperatures $T_j$ and $T_k$ at time $t$. Close to thermal equilibrium the RHS of this equation can be expressed in terms of the radiative thermal conductance 
\begin{equation}
	G=\lim_{\Delta T\rightarrow 0}\frac{\phi^{j\rightarrow k}[T_k(t)+\Delta T,T_k(t)]}{\Delta T},
\end{equation}	
where $\phi^{j\rightarrow k}$ is the net flux (per surface unit) received by the film $k$. This allows us to recast Eq.~\eqref{Eq:SystPvH} under the form
\begin{equation}
		\rho CL \frac{dT_k}{dt} = -G[T_k(t)-T_j(t)].\\
		\label{Eq:SystPvH2}
\end{equation}
If we assume (see Fig.~\ref{Fig:Syst}) that body 2 is in contact with a thermostat at constant temperature $T_2$, the only temperature varying in time is $T_1$, which simply evolves as
\begin{equation}
	T_1(t) = T_1(0) \exp(-t/\tau).
\end{equation}
where $T_1(0)$ is the initial value of the temperature of slab 1, and $\tau=\rho C L/G$ denotes the relaxation time. Hence we see that one way to slow down the thermalization process consists in reducing the coupling strength between the two films, encoded in the conductance $G$. In the following, we show that a strong slowing down of relaxation process can also be observed in some situations when the energy exchange between the two films is strongly coupled to the conduction mechanism within each one of them. 

To demonstrate this result let us first consider the relaxation of two polar films separated by a gap of nanometric thickness. At such separation distance the films interact only by radiation through the tunneling of evanescent photons~\cite{Polder1973}. The radiative power density $P_r^{j\rightarrow k}$ dissipated in the film $k$ at point $z$ and associated with the sources in the other body can be calculated from the average monochromatic flux of the Poynting vector at this point $\langle\mathbf{S}^k(z,\omega)\rangle=2\,\text{Re}\langle \mathbf{E}^k(z,\omega)\times \mathbf{H}^{k*}(z,\omega)\rangle$ as
\begin{equation}
	P_r^{j\rightarrow k}(z)=-\int_0^{\infty} d\omega\,\nabla\cdot\langle \mathbf{S}^{k}(z,\omega)\rangle,
	\label{Power_rad}
\end{equation}
where $\langle . \rangle$ denotes the statistical averaging. According to fluctuational-electrodynamics theory~\cite{RytovBook1989}, for isotropic media and neglecting non-local effects, the Poynting vector reads 
\begin{equation}
	\begin{split}
		\langle S^k_{n}&(z,\omega)\rangle=i\frac{\omega^2}{c^2} \eta_{njl}\\
		&\,\times\int_{\text{sources}}dz' \epsilon''(z',\omega)\Theta[T(z'),\omega]\mathds{G}^{EE}_{j,l}\mathds{G}^{HE*}_{n,l},
		\label{Poynting}
	\end{split}
\end{equation}
where the integral extends over all sources points. In Eq.~\eqref{Poynting}, $\eta_{njl}$ denotes the ${njl}$ component of Levi-Civita tensor , $\Theta(T,\omega)=\hbar\omega/[e^\frac{\hbar\omega}{k_BT}- 1]$ is the mean energy of a Planck oscillator at temperature $T$, $\epsilon''$ the imaginary part of the permittivity in the emitting body while $\mathds{G}^{EE}=\mathds{G}^{EE}(z,z')$ and $\mathds{G}^{HE}=\mathds{G}^{HE}(z,z')$ are the full electric-electric and electric-magnetic dyadic Green tensors~\cite{Tomas} at frequency $\omega$, taking into account all scattering events within the system between the emitter and the point where energy is dissipated. 

As recently established~\cite{Reina2020} the radiative power dissipated through a polar film from its surface is typically reduced by one order of magnitude through a distance of few nanometers from the vacuum gap, so that the radiative transfer can reasonably be assumed to be purely surfacic. In this case, by assuming the thermal conductivity independent of the position and the temperature, and by introducing the auxiliary functions \linebreak $F_k(z,t) = T_k(z,t) - T_2$, the energy balance equation can be recast into the form
\begin{equation}
	\rho C L \frac{\partial F_k(z,t)}{\partial t} = -\kappa \:L\frac{\partial^2 F_k(z,t)}{\partial z^2}+G_r\:F_k(z,t),
	\label{Eq:syst}
\end{equation}
where in this scenario the conductance between the two films reduces to the radiative conductance $G_r$ defined as
\begin{equation}
		G_r =\int\limits_0^{\infty}\frac{\mathrm d\omega}{2\pi}\frac{d\Theta}{dT}(\omega,T)\sum_p \int\limits_0^{\infty}\frac{\mathrm d\kappa}{2\pi} \kappa\,\mathcal T_p(\omega,\kappa,d),
\label{cond_r}
\end{equation}
where $\kappa$ is the modulus of the component of the wavector parallel to the exchange surface and $p$ is the state of polarization. Here $\mathcal T_p(\omega,\kappa,d)$ denotes the energy transmission coefficient for the mode $(\omega,\kappa)$ in polarization $p$ between the films, which can be expressed in terms of reflection and transmission coefficients $r_{ip}$ and $t_{ip}$ of the two slabs as~\cite{Polder1973}
\begin{equation}
	\mathcal T_p(\omega,k,d)=\begin{cases}
		\frac{(1-|r_{1p}|^2-|t_{1p}|^2)(1-|r_{2p}|^2-|t_{2p}|^2)}{|D_p|^2}, & c\kappa< \omega,\\
		\frac{4\,\text{Im}(r_{1p})\text{Im}(r_{2p})e^{-2|k_z|d}}{|D_p|^2}, & c\kappa > \omega,
	\end{cases}
\end{equation}
$k_z$ being the $z$ component of the wavevector and\linebreak $D_p=1-r_{1p}r_{2p}e^{2ik_zd}$ the Fabry-P\'{e}rot factor of the cavity.

As for the initial conditions, we impose \linebreak $F_1(z,0) = \Delta T$ and $F_2(z,0) = 0$, which correspond to the fact the initial temperature profiles in the two slabs are uniform [$T_1(z,0)=T_2+\Delta T$ and $T_2(z,0)=T_2$]. Concerning the boundary conditions, we set $F_2(L+d,t) = 0$ fixing the temperature at $T_2$ for the edge of the right slab in contact with the thermostat, while $\partial_z F_1(-L,t)=0$ imposing a vanishing flux at each instant at the left end of the first slab (adiabatic boundary condition). Notice that this condition is based on the fact that the interactions in the far-field regime with the bath are negligible when the distance $d$ between the slabs is in the near-field regime. Moreover, we impose the two further boundary conditions $\partial_z F_1(0,t) = - G_r/\kappa[F_1(0,t) - F_2(d,t)]$ and $\partial_z F_2(d,t) = -G_r/\kappa[F_1(0,t) - F_2(d,t)]$ ensuring the flux continuity between the two slabs. The solution of the partial differential equations \eqref{Eq:syst} reads
\begin{equation}
	\begin{split}
		F_1(z,t) &= 8 \Delta T \sum_{n=1}^\infty \frac{\sin x_n \cos^2 x_n}{4 x_n + \sin(4 x_n)}	\\
		&\times\cos\Bigl[\frac{x_n(z+L)}{L}\Bigr]\exp\Bigl[-\frac{x_n^2\kappa}{\rho C L^2}t\Bigr],\\
		F_2(z,t) &= - 8 \Delta T \sum_{n=1}^\infty \frac{\sin^2 x_n \cos x_n}{4 x_n + \sin(4 x_n)} \\
		&\times\sin\Bigl(\frac{x_nz}{L}\Bigr)\exp\Bigl[-\frac{x_n^2\kappa}{\rho C L^2}t\Bigr],
	\end{split}
	\label{Eq:T}
\end{equation}
where $x_n$ are the solutions of the transcendental equation $x\tan 2x = G_rL/\kappa$. We can associate each term with a partial relaxation time $\tau_n=\rho C L^2/(x^2_n\kappa)$. It can be easily shown that $x_1<x_n$ and thus $\tau_1>\tau_n$ for all $n\geq2$, so that the first term in these series is the dominant one for large $t$.
\begin{figure}
	\centering
	\includegraphics[width=0.49\textwidth]{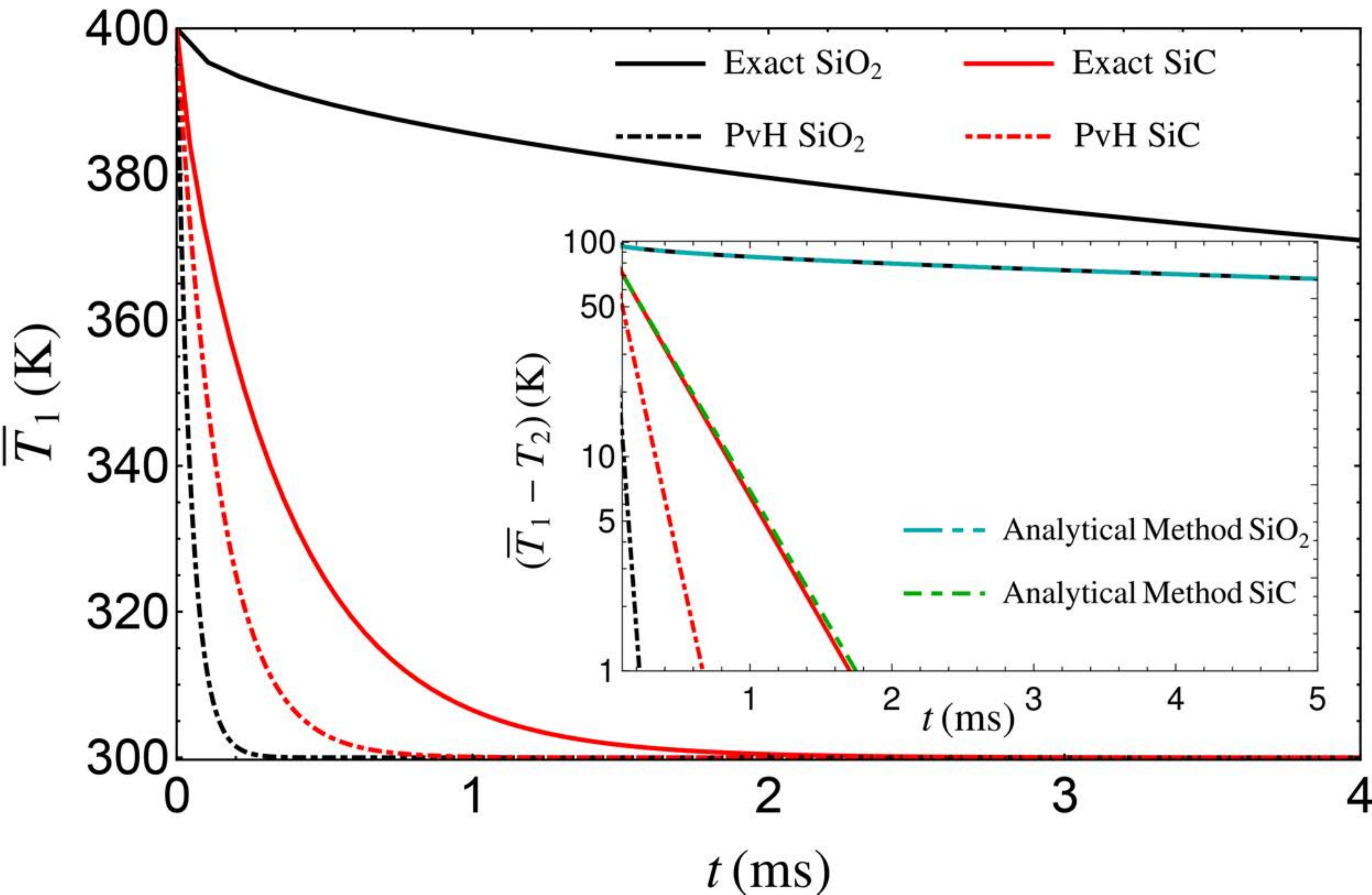}
	\caption{Time evolution of the average temperature of the hot film in a system made of two coupled polar films of thickness $L=100\,\mu$m separated by a vacuum gap of thickness $d=1\,$nm. The black (red) solid curve shows the evolution for SiO$_2$ (SiC) coupled films. The black (red) dashed curves show the evolution predicted by the PvH theory (perfectly conducting solids). The intial temperature of the cold film is $T_2=300\,$K and the initial temperature difference is $\Delta T=100\,$K. The mass density, the specific heat and the thermal conductivity of SiC and SiO$_2$ are $\rho_\text{SiC}=3200\,$kg$\cdot$m$^{-3}$, $C_{\text{SiC}}=600\,$J$\cdot$kg$^{-1}\cdot$K$^{-1}$, $\rho_{\text{SiO}_2}=2650\,$kg$\cdot$m$^{-3}$, $C_{\text{SiO}_2}=680\,$J$\cdot$kg$^{-1}\cdot$K$^{-1}$, $\kappa_{\text{SiC}}=120\,$W$\cdot$m$^{-1}\cdot$K$^{-1}$ and \linebreak $\kappa_{\text{SiO}_2}=1.2\,$W$\cdot$m$^{-1}\cdot$K$^{-1}$.}
	\label{Fig:Tmedio}
\end{figure}
The ratio $G L/\kappa$ quantifies the relative importance of the radiative and conductive transport [for two silicon carbide (SiC) films this ratio is 1.1, while it grows up to 312.5 for two silica (SiO$_2$) films]. The appearance of the ratio $G L/\kappa$ as a key parameters allows us to anticipate a reduction of the effect when reducing the slab thickness $L$ or when increasing the separation distance $d$ (coupling strength) between them~\cite{SupplMat}.

When the heat transport by conduction is much more efficient than the transport by radiation, the temperature within each film is almost uniform at any time. In this case the solution of Eq.~\eqref{Eq:syst} is similar to that of Eq.~\eqref{Eq:SystPvH2} and the temperature profile is the same as the one predicted by the Polder and van Hove (PvH) theory of radiative heat transfer between perfectly conducting solids. However, the situation radically changes when the magnitude of radiative heat transfer is comparable or even larger than the heat transfer by conduction within the films. In Fig.~\ref{Fig:Tmedio} we compare the time evolution of the mean temperature $\overline{T}_1(t)=(1/L)\int T_1(z,t)dz$ inside the left (hot) slab obtained by solving Eq.~\eqref{Eq:Coupling} by means of a finite-difference method~\cite{SupplMat} to the predictions from the PvH theory for SiC and SiO$_2$ films. It can be noticed that the deviation between the two temperatures profiles is more pronounced for two SiO$_2$ slabs compared to SiC slabs. As anticipated previously, this is due to the fact that the thermal conductivities of SiC and SiO$_2$ samples are strongly different. The relatively small conductivity of SiO$_2$ leads to a strong deviation from the PvH predictions. Moreover, as shown in the inset of Fig.~\ref{Fig:Tmedio}, we see that the time evolution of $\overline {T}_1-T_2\,$ is exponentially decaying as predicted by the analytical solution \eqref{Eq:T} and the decay rate of the temperature corresponds to the relaxation time $\tau_1$. Notice also that the comparaison in the inset of the solution of Eq.~\eqref{Eq:syst} with the exact solution of Eq.~\eqref{Eq:Coupling} obtained using a finite-difference method demontrates that the near-field radiative transfer is indeed a surface phenomenon. As shown in the inset of Fig.~\ref{Fig:Tmedio}, the relaxation dynamics is more than one order of magnitude slower when the near-field heat transfer is comparable to the conductive transfer inside the films, showing that the coupling acts as an additional source of thermal inertia. 

\begin{figure}
	\centering
	\includegraphics[width=0.482\textwidth]{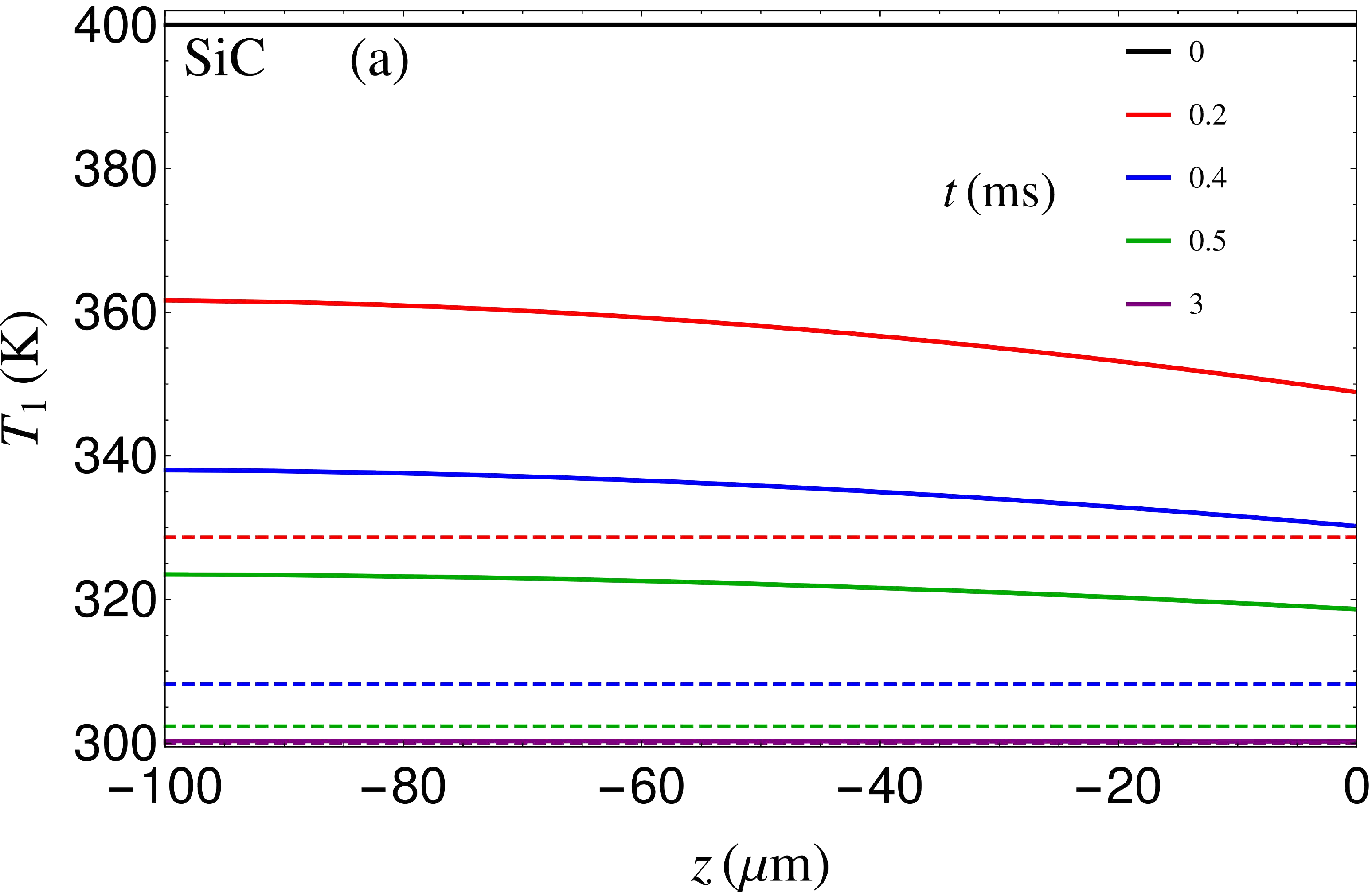}\\
	\includegraphics[width=0.479\textwidth]{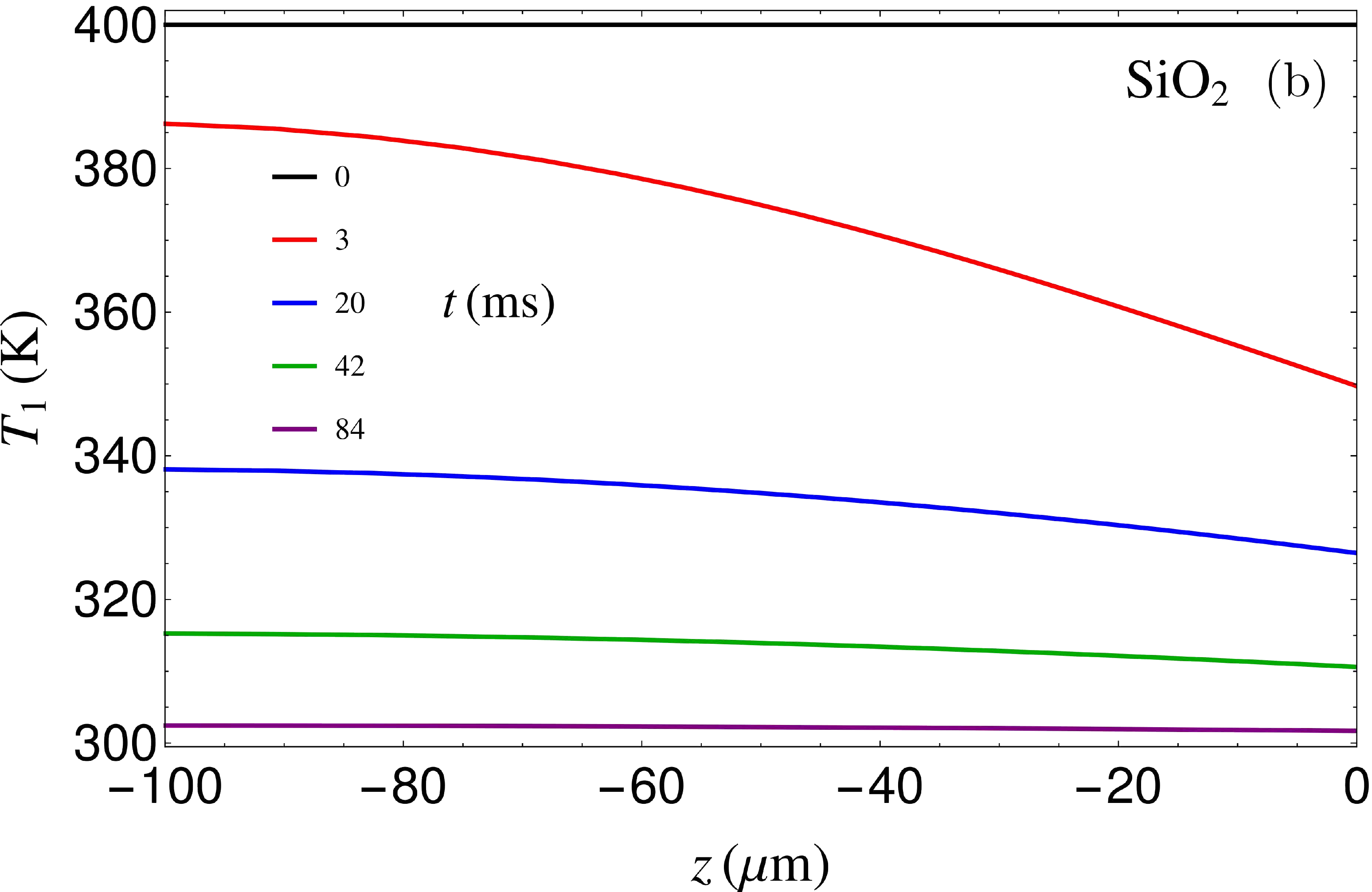}
	\caption{Temperature profile as a function of time in the hot film of a system made of two coupled polar films of thickness $L=100\,\mu$m. The dashed lines correspond to the temperature predicted by the PvH theory. \textbf{(a)} SiC films separated by a $1\,$nm-thick vacuum gap. \textbf{(b)} Same configuration with SiO$_2$ films.}
	\label{Fig:T_z}
\end{figure}

As far as the temperature profile is concerned, we see in Fig.~\ref{Fig:T_z} that it differs from the uniform profile especially when the radiative transfer dominates with respect to the conductive transport, as in the configuration of coupled SiO$_2$ films. As we can see, a non-negligible temperature profile appears through the slab because of the diffusion process in both materials. Althougth the heat spreads inside the films by conduction, the temperature profiles are, of course, not linear because of the interplay between the diffusion process and the near-field heat transfer. For SiC films we also show (dashed lines) the value $T_\text{PvH}$ of the (uniform) temperature predicted by the PvH model at the same moments, while for SiO$_2$ this comparison has been omitted since in this case the two relaxation processes take place at two very different time scales, as shown in Fig.~\ref{Fig:Tmedio}. Indeed in this case, after only $0.2\,$ms the hot film is already thermalized according to the predictions of PvH theory, while a hundred-fold larger time is required when the near-field radiative heat exchange is competing with the diffusion process.

\begin{figure}
	\centering
	\includegraphics[width=0.48\textwidth]{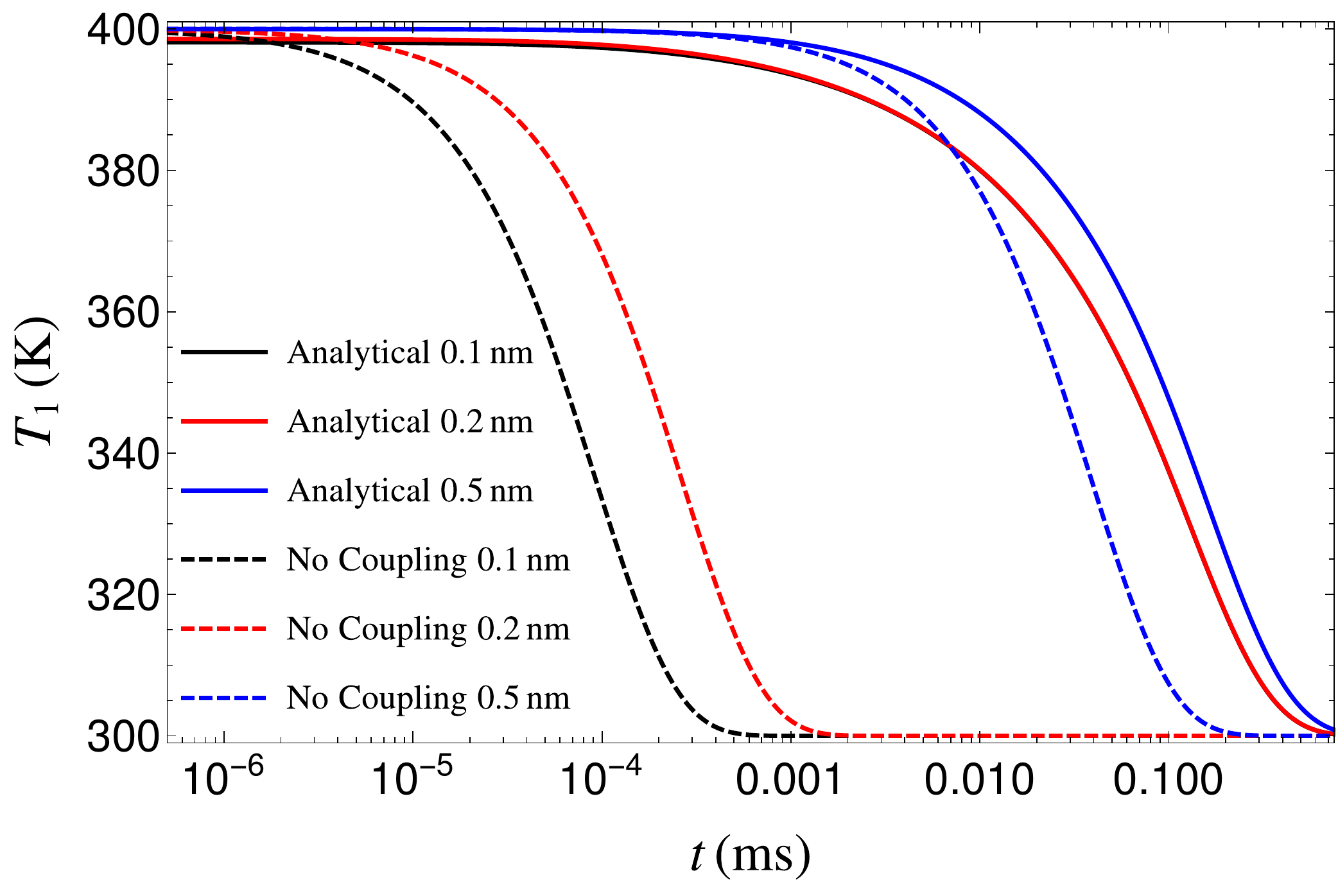}
	\caption{Time evolution of the average temperature of the hot film in a system made of two coupled gold films of thickness $L=100\,\mu$m at subnanometric distances and comparison with the temperature evolution (dashed curves) without coupling. The intial temperature of the cold film is $T_2=120\,$K and the initial temperature difference is $\Delta T=160\,$K. The gold mass density is $\rho=19300\,$kg$\cdot$m$^{-3}$ and its specific heat capacity is $C=128\,$J$\cdot$kg$^{-1}\cdot$K$^{-1}$ .}
	\label{Fig:Au}
\end{figure}

So far we limited ourselves to a transfer between the two solids mediated by photon tunneling. We now consider the thermalization process in the case of two metallic films interacting through electrons tunneling. In fact, it has been recently shown~\cite{messina_heat_2018} that in this scenario the flux carried by electrons at subnanometric distances surpasses by several orders of magnitude the flux carried by photons and can therefore surpass the conductive (phononic) flux inside the metals. The electronic thermal conductance due to electron tunneling can be easily calculated using the effective potential barrier associated with the vacuum gap between the two metals. For two identical metals without bias voltage applied through the system the effective potential is a simple rectangular barrier and the transmission probability $\mathcal{T}(E_z,d)$ at distance $d$ of electrons of normal energy $E_z$ through this barrier reads~\cite{Tannoudji}
\begin{equation}
\mathcal{T}(E_z,d)= \frac{4E_z(E_z-V)}{4E_z(E_z-V)+V^2\sin^2\bigl(k_{2z}(E_z,V)d\bigr)},
\label{Landauer_elec_coeff}
\end{equation}
where $k_{2z}(E_z,V)=\sqrt{2 m_e(E_z-V)}/\hbar$ denotes the normal components of wavector inside the gap ($m_e$ being the electron mass) and the barrier height is written here as $V(d)=V_\text{eV}(d)+E_F$, $E_F$ being the Fermi energy ($E_F=5.53\,$eV for gold) and $V_\text{eV}$ a distant dependent function for which the data taken from \cite{Kiejna} have been fitted from DFT calculations with the log-scale law $V_\text{eV}(d)=V_0 \ln(1+d/1\text{\AA})$ ($V_0=1.25\,$eV for gold). It follows that the heat flux carried by electrons by tunneling effect can be calculated by summing over all energies $E_z$ in the direction normal to the surface. This allows us to define the electronic heat conductance as
\begin{equation}
G_e=\int_{0}^{\infty}dE_z E_z\frac{\partial N(E_z,T)}{\partial T} \mathcal{T}(E_z,d),
\label{elec_flux}
\end{equation}
where $N(E_z,T)dE_z$, with $N(E_z,T)=\frac{m_e k_B T}{2\pi^2\hbar^3}\ln[1+\exp(-(E_z-E_F)/k_B T)]$, denotes the number of electrons in the metal at temperature $T$ with a normal energy between $E_z$ and $E_z+dE_z$ across a unit area per unit time.

This conductance can reach values about six orders of magnitude larger than $G_r$ for gold films at separation distances of few Angstroms~\cite{messina_heat_2018}. In the presence of electron tunneling, the energy-balance equation \eqref{Eq:syst} remains valid, provided that the permutation $G_r\leftrightarrow G_e$ is made. For two gold films, the ratio $G_e L/\kappa$ equals 2.1, 306.5 and 871 for separation distances $d=5\text{\AA}$, $d=2\text{\AA}$ and $d=1\text{\AA}$, respectively. This suggests a strong effect of coupling mechanism on the relaxation dynamics. In Fig.~\ref{Fig:Au} we compare the time evolution of the average temperature profile for the hot film in a system of two gold films with and without coupling between heat conduction and heat transfer by electron tunneling. Unlike polar films interacting by radiation, here the relaxation process toward thermal equilibrium is much faster by two orders of magnitude. But more interesting is the impact of coupling on the relaxation time. At a distance $d=5\text{\AA}$ ($G_e L/\kappa=2.1$) this difference between the scenarios with and without coupling is relatively modest and the coupling slows down the thermalization process by approximatly a factor 2. On the other hand, at closer separation distances this slowdown becomes remarkable, reaching about three orders of magnitude. Notice that for these distances the temporal evolutions of the average temperatures are almost indistinguishable (red and black solid curves) since the $x_n$ in the solution \eqref{Eq:T} of the energy-balance equation (\ref{Eq:syst}) are really close to the asymptotic value (i.e. large value of $G_e L/\kappa$).

In conclusion we have demonstrated a strong impact of the interplay between the heat transfer in extreme near-field regime between two solids and the heat spreading mechanism by conduction inside these media. When the thermal conductance of heat exchange through the separation gap is comparable or dominates over the conductance associated with the diffusive process inside the solids, the thermalization of these media is strongly slowed down. In this case the relaxation time can be larger even by several orders of magnitude than in the classical situation where conduction is the dominant mechanism. This effect should play an important role in the fields of active thermal management at nanoscale, pyroelectric energy conversion in extreme near-field regime, nanoscale heat-engines or for the Boolean treatment of information with heat at nanoscale.

\begin{acknowledgements}
P. B.-A. acknowledges discussions with S.-A. Biehs from Carl von Ossietzky Universität, Germany.

\end{acknowledgements}


\begin{thebibliography}{99}
\bibitem{Evans} D.~J. Evans, D.~J. Searles, and S.~R. Williams, \emph{Dissipation and the relaxation to equilibrium}, J. Stat. Mech.: Theory Exp. P07029 (2009).
\bibitem{Schuster} H.~G. Schuster, \emph{Deterministic Chaos: An Introduction} (Wiley, New York, 1995).
\bibitem{Prigogine} I. Prigogine, \emph{From Being to Becoming: Time and Complexity in the Physical Sciences}, (Freeman, San Franciscom, 1980).
\bibitem{Jin} F. Jin et al., \emph{Equilibration and thermalization of classical systems}, New J. Phys. \textbf{15}, 033009 (2013).
\bibitem{Reid} J.~C. Reid, D.~J. Evans, and D.~J. Searles, \emph{Beyond Boltzmann's H-theorem: demonstration of the relaxation theorem for a non-monotonic approach to equilibrium}, J. Chem. Phys. \textbf{136}, 021101 (2012).
\bibitem{MessinaPRB2013} R. Messina, M. Tschikin, S.-A.Biehs, and P. Ben-Abdallah, \emph{Fluctuation-electrodynamic theory and dynamics of heat transfer in systems of multiple dipoles}, Phys. Rev. B \textbf{88}, 104307 (2013).
\bibitem{superdiff1} P. Ben-Abdallah, R. Messina, S.-A. Biehs, M. Tschikin, K. Joulain, and C. Henkel, \emph{Heat Superdiffusion in Plasmonic Nanostructure Networks}, Phys. Rev. Lett. \textbf{111}, 174301 (2013).
\bibitem{superdiff2} I.Latella et al., \emph{Ballistic near-field heat transport in dense many-body systems}, Phys. Rev. B \textbf{97}, 035423 (2018).
\bibitem{eigen} S. Sanders et al., \emph{Near-Field Radiative Heat Transfer Eigenmodes}, Phys. Rev. Lett. \textbf{126}, 193601 (2021).
\bibitem{phonon} S. Sadasivam, M.~K.~Y. Chan, and P. Darancet, \emph{Theory of Thermal Relaxation of Electrons in Semiconductors}, Phys. Rev. Lett. \textbf{119}, 136602 (2017).
\bibitem{Guha} B. Guha et al., \emph{Near-Field Radiative Cooling of Nanostructures}, Nano Lett. \textbf{12}, 4546 (2012).
\bibitem{logic} P. Ben-Abdallah and S.-A. Biehs, \emph{Towards Boolean operations with thermal photons}, Phys. Rev. B \textbf{94}, 241401(R) (2016).
\bibitem{refrige} K. Chen, P. Santhanam, and S. Fan, \emph{Near-Field Enhanced Negative Luminescent Refrigeration}, Phys. Rev. Appl. \textbf{6}, 024014 (2016).
\bibitem{Yu} R. Yu, A. Manjavacas, and F.~J. Garc\'{i}a de Abajo, \emph{Ultrafast radiative heat transfer}, Nat. Commun. \textbf{8}, 1 (2017).
\bibitem{Transi} I. Latella et al., \emph{Dynamical Response of a Radiative Thermal Transistor Based on Suspended Insulator-Metal-Transition Membranes},
Phys. Rev. Appl. \textbf{11}, 024004 (2019).
\bibitem{Bhatia} B. Bhatia, H. Cho, J. Karthik, J. Choi, D.~G. Cahill, L.~W. Martin and W.~P. King, \emph{High Power Density Pyroelectric Energy Conversion in Nanometer-Thick BaTiO$_3$ Films}, Nanoscale and Microscale Thermophys. Eng. \textbf{20}, 137 (2016).
\bibitem{Kubo} R. Kubo, M. Toda, and N. Hashitsume, \emph{Statistical Physics II: Nonequilibrium Statistical Mechanics}, vol 31, Springer Series in Solid–State Science, (Springer, New York, 1985).
\bibitem{Chen} V. Chiloyan, J. Garg, K. Esfarjani, and G. Chen, \emph{Transition from near-field thermal radiation to phonon heat conduction at sub-nanometre gaps}, Nat. Commun. \textbf{6}, 6755 (2015).
\bibitem{Reddy} K. Kim et al.,\emph{Radiative heat transfer in the extreme near field}, Nature \textbf{528}, 387 (2015).
\bibitem{Kittel} K. Kloppstech et al.,\emph{Giant heat transfer in the crossover regime between conduction and radiation}, Nat. Commun. \textbf{8}, 14475 (2017).
\bibitem{Messina2016} R. Messina, W. Jin, and A.~W. Rodriguez, \emph{Strongly coupled near-field radiative and conductive heat transfer between planar bodies}, Phys. Rev. B \textbf{94}, 121410(R) (2016).
\bibitem{Reina2020} M. Reina, R. Messina, and P. Ben-Abdallah, \emph{Conduction-radiation coupling between two closely-separated solids}, Phys. Rev. Lett. \textbf{125}, 224302 (2021).
\bibitem{Polder1973} D. Polder and M. van Hove, \emph{Theory of Radiative Heat Transfer between Closely Spaced Bodies}, Phys. Rev. B \textbf{4}, 3303 (1971).
\bibitem{RytovBook1989} S.~M. Rytov, Y.~A. Kravtsov, and V.~I. Tatarskii, \emph{Principles of Statistical Radiophysics}, Vol. 3 (Springer, New York, 1989).
\bibitem{Tomas} M.~S. Toma\v{s}, \emph{Green function for multilayers: Light scattering in planar cavities}, Phys. Rev. A \textbf{51}, 2545 (1995).
\bibitem{SupplMat} See EPAPS Document No. [number will be inserted by publisher] for a description of the finite-difference method used to solve Eq.~\eqref{Eq:Coupling} and of the dependence of the mean temperature of films on their separation distance and on their thickness.
\bibitem{messina_heat_2018} R.~Messina, S.-A.~Biehs, T.~Ziehm, A.~Kittel, and P.~Ben-Abdallah, \textit{Heat transfer between two metals through subnanometric vacuum gaps}, arXiv:1810.02628 (2018).
\bibitem{Tannoudji} C. Cohen-Tannoudji, J. Dupont-Roc, and G. Grynberg, \emph{Atom-photon interactions} (Wiley, New York, 1992).
\bibitem{Kiejna} A. Kiejna, \emph{Potential barrier for the metal-vacuum-metal tunneling electrons}, Ultramicroscopy \textbf{42}, 231 (1992).
\bibitem{Zhang} P. Zhang, \emph{Scaling for quantum tunneling current in nano- and subnano-scale plasmonic junctions}, Sci. Rep. \textbf{5}, 9826 (2015).
\bibitem{Marschall} J. Marschall and A. Majumdar, \emph{Charge and energy transport by tunneling thermoelectric effect}, J. Appl. Phys. \textbf{74}, 4000 (1993).
\end{thebibliography}
\end{document}